%% file: late14ae.tex
\documentclass[a4paper,fleqn,usenatbib]{mnras}
\usepackage{footmisc}
\usepackage{graphicx}
\usepackage{amssymb}
\usepackage{amsmath}
\usepackage{wasysym}
\usepackage{color}
\usepackage[normalem]{ulem}
\usepackage{psfrag}
\usepackage{multirow,bigdelim}
\graphicspath{{./}{../plots/}}

\newcommand{\halpha}{\hbox{H$\alpha$}}
\newcommand{\hbeta}{\hbox{H$\beta$}}

\newcommand{\heii}{\hbox{\ion{He}{ii}~$\lambda 4686$}}

\newcommand{\asae}{\hbox{ASASSN-14ae}}

\newcommand{\mbh}{\hbox{$M_{\rm BH}$}}
\newcommand{\msun}{M_{\odot}}
\newcommand{\rsol}{R_{\odot}}

\newcommand{\SWIFT}{\textit{Swift}}
 
\newcommand{\IRAF}{\textsc{iraf}}

\newcounter{minirefcount}
\title[The Fading of \asae]{Hello Darkness My Old Friend: The Fading of the Nearby TDE ASASSN-14ae}

\author[J. S. Brown et al.]{Jonathan S. Brown,$^{1}$\thanks{E-mail: brown@astronomy.ohio-state.edu}
Benjamin J. Shappee,$^{2,3,4}$
T. W.-S. Holoien,$^{1,5}$
K. Z. Stanek,$^{1,5}$ \newauthor
C. S. Kochanek,$^{1,5}$ 
and J. L. Prieto$^{6,7}$  
\\
$^{1}$ Department of Astronomy, The Ohio State University, 140 West 18th Avenue, Columbus, OH 43210, USA\\
$^{2}$ Carnegie Observatories, 813 Santa Barbara Street, Pasadena, CA 91101, USA\\
$^{3}$ Hubble Fellow\\
$^{4}$ Carnegie-Princeton Fellow\\
$^{5}$ Center for Cosmology and Astro-Particle Physics, The Ohio State University, 191 West Woodruff Avenue, Columbus, OH 43210, USA\\
$^{6}$ N\'ucleo de Astronom\'ia de la Facultad de Ingenier\'ia, Universidad Diego Portales, Av. Ej\'ercito 441, Santiago, Chile \\
$^{7}$ Millennium Institute of Astrophysics, Santiago, Chile \\
}

\date{Accepted XXX. Received YYY; in original form ZZZ}

\pubyear{2016}

\begin{document}
\label{firstpage}
\pagerange{\pageref{firstpage}--\pageref{lastpage}}
\maketitle

\begin{abstract}
We present late-time optical spectroscopy taken with the Large Binocular Telescope's Multi-Object Double Spectrograph, an improved ASAS-SN pre-discovery non-detection, and late-time \SWIFT\ observations of the nearby ($d=193$ Mpc, $z=0.0436$) tidal disruption event (TDE) \asae. Our observations span from $\sim$~20 days before to $\sim$~750 days after discovery. The proximity of \asae\ allows us to study the optical evolution of the flare and the transition to a host dominated state with exceptionally high precision. We measure very weak \halpha\ emission 300 days after discovery ($L_{\rm H\alpha} \simeq 4\times 10^{39}$ ergs s$^{-1}$) and the most stringent upper limit to date on the \halpha\ luminosity $\sim$~750 days after discovery ($L_{\rm H\alpha} \lesssim 10^{39}$ ergs s$^{-1}$), suggesting that the optical emission arising from a TDE can vanish on a timescale as short as 1 year. Our results have important implications for both spectroscopic detection of TDE candidates at late times, as well as the nature of TDE host galaxies themselves. 
\end{abstract}


\section{Introduction}
\label{sec:intro}

Tidal disruption events (TDEs) occur when a star's self gravity is overcome by the tidal forces from a nearby supermassive black hole (SMBH). The characteristic radius at which a star is disrupted is approximately $r_p \sim R_*(\mbh/M_*)^{1/3}$. For a main-sequence star, approximately half of the stellar debris will remain on bound orbits and return to pericenter at a rate proportional to $t^{-5/3}$ \citep{Rees88,Evans89,Phinney89}. For $\mbh \gtrsim 10^8 \msun \left(R_*/\rsol \right)^{3/2} \left(M_*/\msun \right)^{-1/2}$, $r_p$ falls roughly within the Schwarzschild radius and the star is simply absorbed. For black holes less massive than $\sim 10^8\msun$, the characteristics of the observed emission are heavily dependent on the disrupted star \citep[e.g.][]{MacLeod12,Kochanek16_stellar}, the post-disruption evolution of the accretion stream \citep[e.g.][]{Kochanek94,Strubbe09,Guillochon13,Piran15,Shiokawa15}, and complex radiative transfer effects \citep[e.g.][]{Gaskell14,Strubbe15,Roth15}, making TDEs exceptional probes of both SMBH physics and environment \citep[e.g.][]{Magorrian99,Ulmer99,Wegg11,Metzger15,Li15,Ricarte15}.

\input{./tab1.tex}

Given the small number of convincing TDE candidates\footnote{https://tde.space/}, many of which have only sparse data, the late-time characteristics of these exotic objects have not been well established. The number of optical TDE candidates is growing, and some previous studies have even examined the spectral characteristics of TDEs both near peak and at later times \citep[e.g.][]{vanVelzen11,Cenko12,Gezari12,Arcavi14,Chornock14,Holoien14,Gezari15,Vinko15,Holoien16_15oi,Holoien16_14li}. However, the majority of these studies were focused on either characterizing the immediate evolution of the flare, or studying the underlying host galaxies. As a result, the spectroscopic follow-up observations were conducted on either very short or very long timescales after the flares occurred, and consequently provided only weak constraints on the late-time spectral evolution. \citet{vanVelzen11} examined the flaring state characteristics of TDE1 and TDE2, but since their study was based primarily on archival data, they were not able to study the spectroscopic evolution of the flares in detail. \citet{Gezari12} presented moderately late-time ($\sim$~250 days) follow-up spectroscopy of the H-deficient TDE PS1-10jh, but broad \heii\ emission was still visible well above the host continuum in their late-time spectrum. In contrast, \citet{Gezari15} presented host galaxy spectra taken years after the flare had faded. Similarly, \citet{Arcavi14} and \citet{French16} presented late-time spectra of several recent TDEs, but in each case the spectrum was either dominated by the flare, or the observations were conducted years after the flare had faded, leaving the transition from a flare dominated state to a host galaxy dominated state relatively unobserved. The spectroscopic characteristics of some peculiar optical transients have been studied during the transition to the host dominated regime (e.g. PTF10iya; \citealt{Cenko12}, PS1-11af; \citealt{Chornock14}, Dougie; \citealt{Vinko15}, and ASASSN-15oi; \citealt{Holoien16_15oi}). With the exception of ASASSN-15oi, these TDE candidates are relatively distant, making the characterization of their spectroscopic evolution difficult. Finally, most studies of TDEs lack a quantitative estimate of the luminosity evolution of various spectral features.

ASASSN-14ae \citep{Prieto14,Holoien14} was a nearby ($d\sim$~200 Mpc, $z=0.0436$) TDE discovered by the All-Sky Automated Survey for SuperNovae \citep[ASAS-SN;][]{Shappee14} on 2014-01-25.51. An immediate follow-up campaign \citep{Holoien14} observed \asae\ for $\sim$~150 days. We found the evolution of \asae\ to be consistent with that of a blackbody with constant temperature and exponentially declining luminosity. Although the relative strength of the \heii\ and Balmer lines appears to vary with time, we also found that the spectral characteristics of \asae\ fell near the middle of the H-to-He dominated continuum proposed by \citet{Arcavi14}.

In this paper we present late-time observations of \asae\ that follow the transition from a flare dominated state to a host galaxy dominated state. We present an improved ASAS-SN pre-discovery upper limit, as well as extensive follow-up data consisting of optical spectra taken with the Multi-Object Double Spectrograph 1 (MODS1) on the 8.4 m Large Binocular Telescope (LBT), and UVOT/X-ray observations from the \SWIFT\ space telescope, which provide an exceptional opportunity for characterizing the late-time evolution of a TDE. Our limits on the previously strong \halpha\ and UV emission are the most stringent limits to date on late-time emission from a TDE. In Section~\ref{sec:data} we describe our observations, in Section~\ref{sec:discussion} we present our measurements of the late-time evolution, and finally in Section~\ref{sec:conclusions} we provide a summary of our results and discuss the implications for future studies.


\section{Observations}
\label{sec:data}

\subsection{Spectroscopic Observations}

Follow-up spectroscopy of \asae\ was obtained with MODS1 \citep{Pogge10} on the LBT between February 2014 and February 2016. Observations were performed in longslit mode with a 1\farcs0 slit, with the exception of the latest epoch, which used a 1\farcs2 slit. MODS1 uses a dichroic that splits the light into separately optimized red and blue channels at $\sim$~5650\,\AA. The blue CCD covers a  wavelength range of $\sim$~3200 -- 5650\,\AA, with a spectral resolution of 2.4\,\AA, while the red CCD covers a wavelength range of $\sim$~5650 -- 10000\,\AA, with a spectral resolution of 3.4\,\AA.

Our first spectrum was taken on 2014-02-24 ($t = 29.7$ days after discovery; \citealt{Prieto14}), and consisted of three 300s exposures. The following two observations ($t = 93.8, 131.7$ days) consisted of three 1050s exposures. These spectra were presented in Figure 5 of \citet{Holoien14} but the detailed analysis of the evolution of \halpha\ was restricted to the first $\sim$~70 days. We obtained two additional observations unique to this paper, on 2014-11-22 (6$\times$1200s) and 2016-02-08 (3$\times$600s) corresponding to 301.0 and 743.5 days after discovery, respectively. The position angle of the slit approximated the parallactic angle at the midpoint of the observations in order to minimize slit losses due to differential atmospheric refraction. We obtained bias frames and Hg(Ar), Ne, Xe, and Kr calibration lamp images for wavelength calibration. If the arc lamp or flat field data were not available on the night of the observation, we used calibration data obtained within 1-2 days of our observations. Given the stability of MODS1 over the course of an observing run, this is sufficient to provide accurate calibrations. Night sky lines were used to correct for the small ($\sim$~1\,\AA) residual flexure. Standard stars were observed with a 5$\times$60\arcsec\ spectrophotometric slit mask and used to calibrate the response curve. The standard stars are from the HST Primary Calibrator list, which is composed of well observed northern-hemisphere standards from the lists of \citet{Oke90} and \citet{Bohlin95}. We list the information regarding our observational configuration in Table~\ref{tab:tab1}.

We used the modsCCDRed\footnote{http://www.astronomy.ohio-state.edu/MODS/Software/modsCCDRed/} suite of \textsc{python} programs to bias subtract, flat field, and illumination correct the raw data frames. We removed cosmic rays with L.A. Cosmic \citep{vanDokkum01}. The sky subtraction and one-dimensional extraction were performed with the modsIDL pipeline\footnote{http://www.astronomy.ohio-state.edu/MODS/Software/modsIDL/}. We correct residual sky features with reduced spectra of standard stars observed on the same night under similar conditions and the \IRAF\ task \textit{telluric}. Finally, the individual spectra from each night were combined, yielding a total of five high S/N spectra corresponding to the five observation epochs.

Our spectra were taken under variable conditions with a relatively narrow (1\arcsec) slit. In order to facilitate comparison of spectra across multiple observing epochs, we calibrated the flux of each spectra with the contemporaneous $r'$-band MODS acquisition images. We performed aperture photometry on the \asae\ host and bright stars in the field with the \IRAF\ package \textit{apphot}. We scaled the magnitudes of the stars to match their SDSS $r'$-band magnitudes, and applied the same scale factor to the host of \asae. Finally, following \citet{Shappee13}, we scale each spectrum of \asae\ such that its synthetic $r'$-band photometry matched the corresponding $r'$-band aperture photometry. The spectral evolution of \asae\ is presented in Section~\ref{sec:spec}.

\subsection{\SWIFT\ Observations}

After the publication of \citet{Holoien14}, we also obtained additional \SWIFT\ observations of \asae. The UVOT \citep{Poole08} observations were obtained in six filters: $V$ (5468 \AA), $B$ (4392 \AA), $U$ (3465 \AA), $UVW1$ (2600 \AA), $UVM2$ (2246 \AA), and $UVW2$ (1928 \AA). We used the UVOT software task \textsc{uvotsource} to extract the source counts from a 5.0\arcsec radius region and a sky region with a radius of $\sim$~40\arcsec. The UVOT count rates were converted into magnitudes and fluxes based on the most recent UVOT calibration \citep{Poole08, Breeveld10}. For the most recent observations, we coadded several exposure taken over a $\sim$~2 week period in order to obtain a high S/N determination of the host magnitudes. 

We also obtained X-ray observations with the \SWIFT\ X-ray Telescope \citep[XRT;][]{Burrows05}. The XRT was operating in Photon Counting mode \citep{Hill04} during all Swift observations. We reduced and combined all 30 epochs of observations using the software tasks {\sc xrtpipeline} and {\sc xrtselect} to produce an image in the $0.3-10$~keV range with a total exposure time of $\sim58000$~s. We extracted source counts and background counts using a region with a radius of 20 pixels (47\farcs{1}) centered on the position of \asae\ and a source-free region with radius of 70 pixels (165\farcs{0}), respectively. The results of our X-ray and UVOT photometric analysis are presented in Section~\ref{sec:phot}.

\subsection{ASAS-SN Pre-Discovery Upper Limit}
To further constrain the early-time light curve of ASASSN-14ae we re-evaluate the pre-discovery ASAS-SN non-detection first reported by \citet{Prieto14}.  The last ASAS-SN epoch before discovery was observed on 2014-01-01.53 under favorable conditions by the quadruple 14-cm ``Brutus'' telescope in Haleakala, Hawaii.  This ASAS-SN field was run through the standard ASAS-SN pipeline (Shappee et al. in prep.) using the \textsc{isis} image subtraction package \citep{Alard98, Alard00}, except we did not allow images acquired between 2013-11-03 and 2015-12-13 to be used in the construction of the reference image to avoid light from \asae\ contaminating the reference. We then performed aperture photometry at the location of ASASSN-14ae on the subtracted images using the \textsc{IRAF} \textit{apphot} package and calibrated the results using the AAVSO Photometric All-Sky Survey \citep[APASS;][]{Henden15}.  There was no excess flux detected at the location of ASASSN-14ae over the reference image on 2014-01-01.53 ($t = -23.98$ days), and we place a 5-sigma limit of $V>18.21$ mag on ASASSN-14ae at this epoch. 


\section{Evolution of the Late-Time Emission}
\label{sec:discussion}

\input{./tab2.tex}

\begin{figure*}
\centering{\includegraphics[scale=1.,width=\textwidth,trim=0.pt 0.pt 0.pt 0.pt,clip]{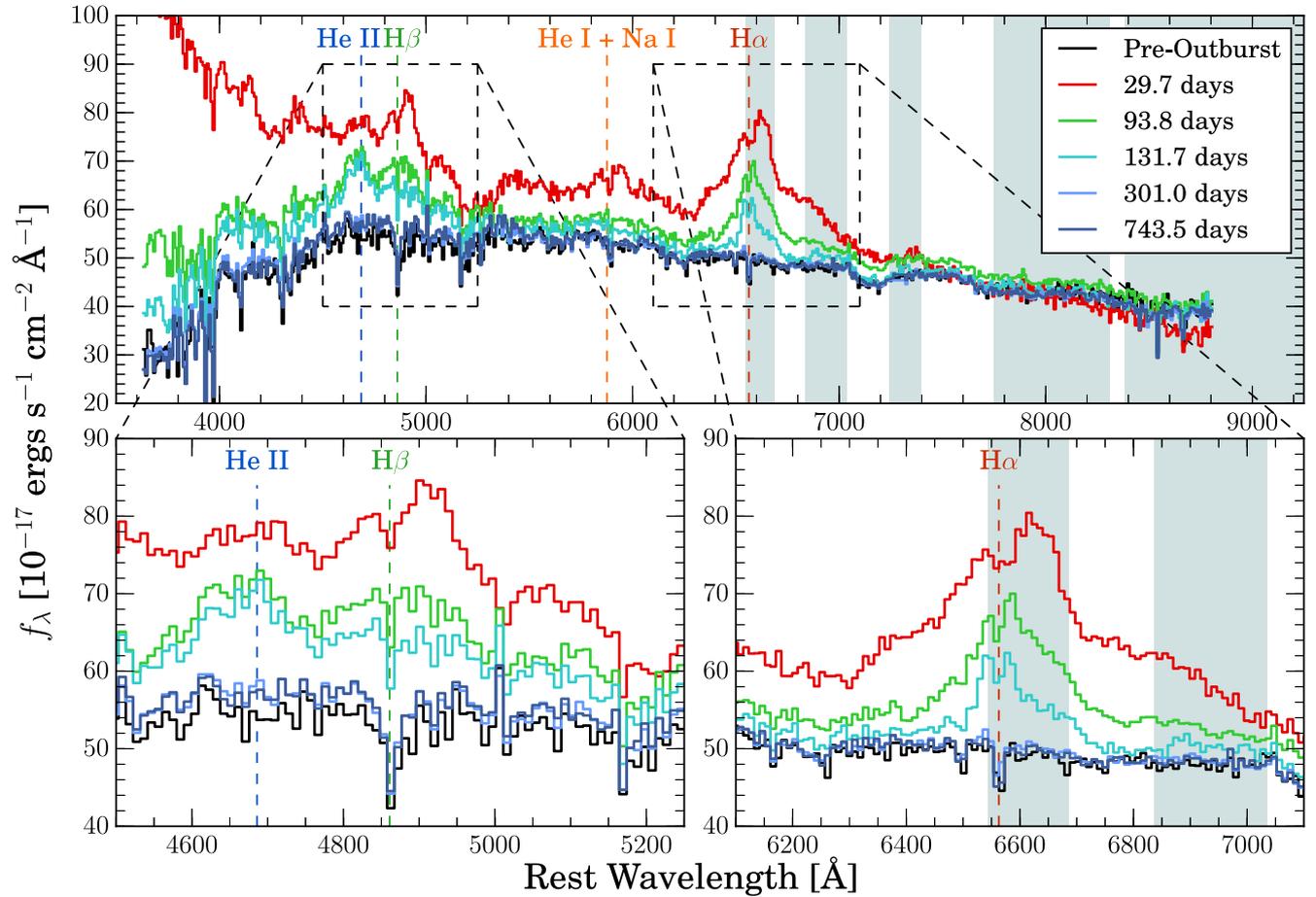}}
\caption{Rest frame absolute-flux-calibrated spectra of \asae. Color denotes days since discovery. The top panel shows the full optical spectrum, while the bottom panel shows zoomed portions of the regions in the immediate vicinity of \heii\ (left) and \halpha\ (right). The shaded regions show the location of telluric features where systematic errors may be significant. The spectra show clear temporal evolution in the sense of decreasing continuum and emission line features with increasing time. We find no evidence for TDE emission in the spectrum taken at $\sim$~750 days, and adopt this as a nominal host spectrum.}
\label{fig:specEvol}
\end{figure*}

\begin{figure}
\centering{\includegraphics[scale=1.,width=0.45\textwidth,trim=0.pt 0.pt 0.pt 0.pt,clip]{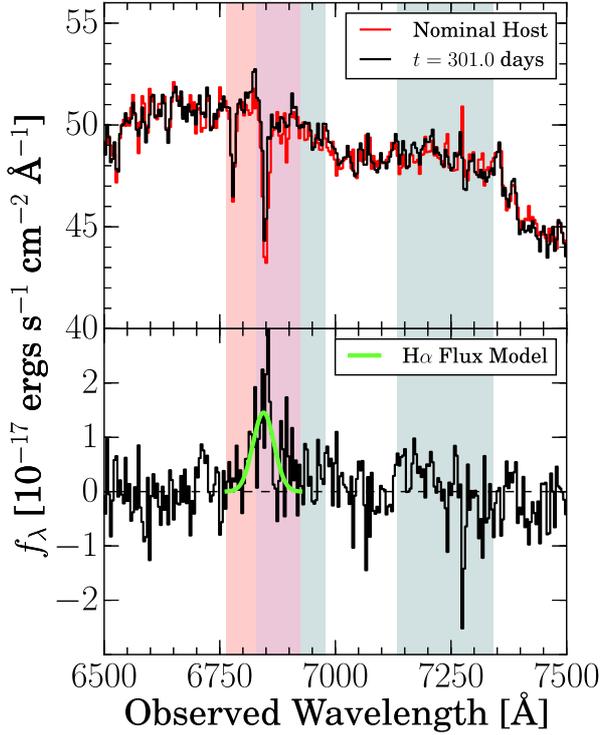}}
\caption{Top: late-time spectra of \asae, showing the spectrum taken at $\sim$~300 days (black) and the host spectrum taken at $\sim$~750 days (red), both binned to the approximate spectral resolution ($\sim$~4\AA). Bottom: residuals after host subtraction. The red shaded band shows the $\sim$~150\AA\ wide region ($\pm3500$ km s$^{-1}$) of the spectrum we searched for \halpha\ emission, while the gray shaded bands shows the regions of the spectrum prone to systematic errors associated with telluric features. The green curve shows our best fit model for the emission we find in the vicinity of \halpha.}
\label{fig:trace}
\end{figure}

\begin{figure}
\centering{\includegraphics[scale=1.,width=0.45\textwidth,trim=0.pt 0.pt 0.pt 0.pt,clip]{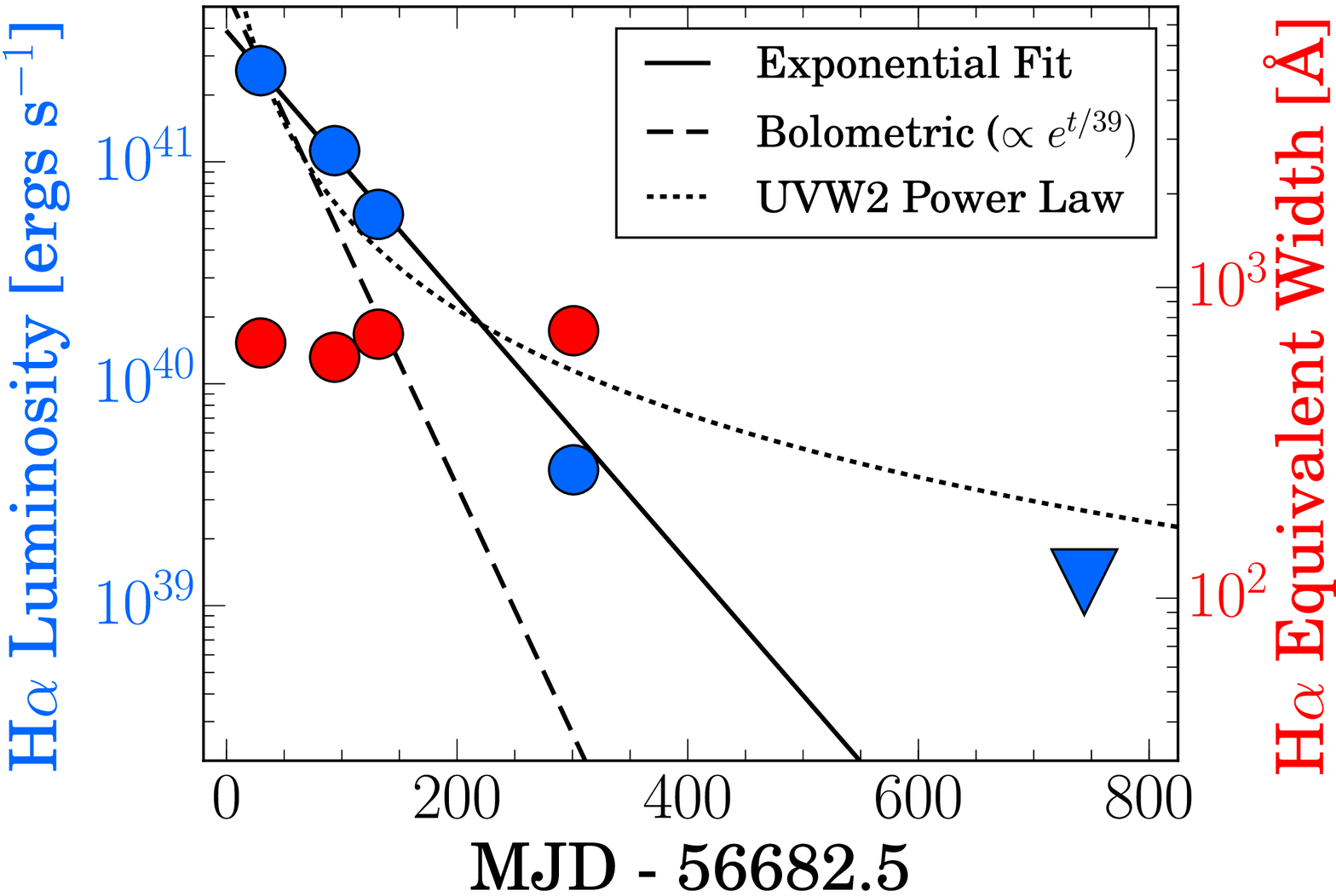}}
\caption{Luminosity and equivalent width evolution of the \halpha\ emission line feature up to $\sim$~750 days after discovery. The left (right) axis shows the measured luminosity (equivalent width) of the \halpha\ emission. Circles show our detections, and triangles denote upper limits. The solid line shows our fit to the \halpha\ luminosity evolution assuming an exponential decay, while the dashed and dotted lines show the expected evolution if the \halpha\ luminosity followed the bolometric evolution from \citet{Holoien14} and a power-law fit to the UVW2 emission, respectively.}
\label{fig:lineEvol}
\end{figure}

\subsection{Spectral Analysis}
\label{sec:spec}

In \citet{Holoien14} we showed that the evolution of \asae\ is consistent with a constant temperature blackbody and exponentially decreasing luminosity. We also analyzed the spectral characteristics up to 70 days after discovery and showed that both the blue continuum and the strength of prominent emission lines (\halpha, \hbeta, and \heii) decreased with time, while the strength of \heii\ relative to the Balmer lines increased. The \halpha\ profile was analyzed up to $\sim70$ days after discovery and was shown to evolve from what was initially a very broad (FWHM $\sim$~20000 km s$^{-1}$) symmetric profile to a slightly narrower (FWHM $\sim$~10000 km s$^{-1}$) asymmetric profile, and then to narrower still (FWHM $\sim$~8000 km s$^{-1}$) symmetric profile. The \halpha\ emission also showed a consistent red offset ($\Delta v\sim$~2000--4000 km s$^{-1}$) from the systemic velocity.

The additional observations presented here provide further constraints on the evolution of the TDE flare. Figure~\ref{fig:specEvol} shows the evolution of the optical emission beginning $\sim$~30 days after discovery (red) and ending $\sim$~750 days after discovery (dark blue). The black spectrum shows the archival SDSS DR7 \citep{Abazajian09} spectrum of the host galaxy taken on 2008-02-17 for a total exposure time of $2220.50$s. The top panel shows the full optical spectrum, while the bottom panels show expanded views of the \heii\ (left) and \halpha\ (right) regions. Prominent spectral features are labeled, while the shaded bands denote regions prone to systematic errors related to telluric correction. In particular, the B-band telluric feature is in close proximity to \halpha\ at the redshift of \asae.

In order to measure the \halpha\ emission from the TDE, we must first subtract the underlying host galaxy. All signatures of the flare have vanished by $\sim$~750 days, which allows us to precisely subtract the host spectrum without introducing uncertainties associated with modeling the underlying population \citep[e.g.][]{Gezari15}. Furthermore, using the host spectrum obtained with MODS, rather than the archival SDSS spectrum, increases our signal to noise in the subtracted spectrum and eliminates systematic errors caused by differences in the spatial coverage of the two instruments, which are likely to be significant. After subtracting the host spectrum, we fit the \halpha\ profile with a Gaussian superimposed on a low-order continuum. We subtract the continuum and measure the line flux by directly integrating the observed \halpha\ profile within $\pm 3\sigma$ of the line center. While the observed \halpha\ profiles are typically asymmetric and more strongly peaked than a Gaussian, this method appears to be accurate to $\sim$~10\%, which is sufficient for our purposes. We list our measurements of the \halpha\ line profile in Table~\ref{tab:tab2}.

Our initial spectrum taken 30 days after discovery (red) shows broad (FWHM $\sim$~13000 km s$^{-1}$) emission lines superimposed on a strong blue continuum. The lines are redshifted with respect to the host galaxy ($\Delta v\sim$~2000 km s$^{-1}$). The optical emission from the flare gradually decays with time and by $\sim$~90 days (green), the continuum emission has weakened substantially along with the \halpha\ and \hbeta\ emission lines. The strength of the \heii\ line has grown to be comparable to that of \hbeta, as noted in previous studies \citep{Holoien14,Arcavi14}. Approximately 130 days after discovery (cyan), there is very little continuum emission remaining, while the \halpha, \hbeta, and \heii\ emission features are still rather prominent. The width of the \halpha\ feature ($\sim$~5000 km s$^{-1}$) has decreased by a factor of $\sim$~2 since our observation $\sim$~30 days after discovery, and maintains a modest red offset from the systemic velocity ($\Delta v\sim$~400 km s$^{-1}$). Our deepest spectrum (blue), taken at $\sim$~300 days, shows only marginal evidence for any optical emission whatsoever, and looks nearly identical to both the spectrum taken at $\sim$~750 days (dark blue) and the pre-outburst SDSS spectrum (black).

Figure~\ref{fig:trace} shows the results of our host subtraction after $\sim$~300 days. The top panel shows the observed spectrum (black) and the estimate of the host spectrum (red), both binned to the approximate spectral resolution ($\sim$4\AA). The bottom panel shows the residuals of our host subtraction. The gray bands show the regions of the spectrum prone to systematic effects associated with the telluric correction, and the red band shows the $\pm3500$~km s$^{-1}$ region where we expect to see residual \halpha\ emission from the TDE. We find excess emission relative to the host spectrum; the green curve shows our best fit to the residual \halpha\ emission assuming the feature is roughly Gaussian.

We find no evidence for excess \halpha\ emission in our latest spectrum relative to the archival SDSS spectrum of the host. The lower left panel of Figure~\ref{fig:specEvol} appears to show evidence for excess blue emission relative to the SDSS spectrum. However, the fact that all wavelengths blueward of the MODS1 dichroic show uniform excess flux relative to the SDSS spectrum suggests that this is likely a systematic effect associated with our flux calibration rather than excess emission due to the TDE. Additionally, we find that the equivalent widths of strong absorption lines (e.g. \hbeta) are consistent between the archival SDSS spectrum and our latest spectrum. 

While our non-detection of residual emission precludes the determination of an upper limit on the \halpha\ equivalent width, we are able to place an upper limit on the \halpha\ luminosity. We measure the RMS of the residuals between the MODS spectrum from $\sim750$~days and the archival SDSS spectrum, and assuming a FWHM of the \halpha\ feature of $\sim 2000$ km s$^{-1}$ and a spectral resolution of $\sim4$\AA\ yields a 1-$\sigma$ upper limit on the \halpha\ luminosity of $1.3\times10^{39}$ ergs s$^{-1}$.

We quantify the evolution of the \halpha\ profile in Figure~\ref{fig:lineEvol}. We show the luminosity (blue points, left axis) and equivalent width (red points, right axis) of the $\halpha$ line as a function of time. The triangle denotes our upper limits on the \halpha\ emission. We also show various curves that describe the luminosity evolution of \asae. The thick solid line shows our fit to the \halpha\ luminosity evolution assuming an exponential decline. Interestingly, the \halpha\ luminosity evolves on a significantly longer timescale than implied by the bolometric luminosity evolution from \citet{Holoien14} (dashed line). In contrast, both exponential profiles decline much faster than typical power-law models \citep{Holoien14}, though we lack the crucial early time data needed for constraining the power-law models. Assuming the \halpha\ emission is driven primarily by photoionization and recombination, we expect the \halpha\ emission to track the shortest wavelength UVW2 evolution more than any other UVOT band. The dotted line shows our best fit to the \halpha\ luminosity assuming $L_{\rm H\alpha} \propto (t-t_0)^{-5/3}$ \citep{Evans89,Phinney89}, where $t_0 = -18.0$ days is based on a fit to the UVW2 early time photometry. While the power law appears to describe the photometry reasonably well (see Figure~\ref{fig:swift}), this model over-predicts the observed \halpha\ luminosity by a factor of $\gtrsim$~2 at late times, even with the most accommodating value of $t_0$ permitted by the ASAS-SN data ($t_0 = -23.98$ days).

While the \halpha\ luminosity decreases with time, the \halpha\ equivalent width increases slightly during the 300 day period following the flare. The measurement of equivalent width is sensitive to seeing variations as well as slit positioning, but our observations are largely consistent with the \halpha\ emission decreasing more slowly than the bolometric luminosity from \citet{Holoien14}. If the \halpha\ emission is driven primarily by photoionization and recombination, this implies that the flux in the far-UV decreases more slowly than the near-UV/optical continuum. Similarly, the increasing strength of \heii\ relative to \hbeta\ supports the hardening of the UV spectrum at late times. This is consistent with some models \citep[e.g.][]{Lodato11,Metzger15}, which predict that at later times the reprocessing envelope dissipates and allows higher energy photons to escape, ultimately giving rise to a harder observed spectrum. However, in contrast with this picture, we detect no signs of X-ray emission (see Section~\ref{sec:phot}). Furthermore, as the luminosity decreases and radiation escapes from regions closer to the black hole, the observed line widths could be expected to increase, similar to what is seen for AGN \citep{McGill08,Denney09}. Interestingly, this is the opposite of what we observe in \asae\ as well as other TDEs \citep{Holoien16_15oi,Holoien16_14li}, in which the line widths appear to decrease with decreasing luminosity. The decreasing line width may suggest that the line emission arises from predominantly larger radii at later times \citep[e.g.][]{Guillochon14}.

Our analysis is not without systematic uncertainties. Figures~\ref{fig:specEvol}~and~\ref{fig:trace} show that at a redshift of $z=0.0436$, the \halpha\ line is in close proximity to the B-band telluric feature. While we have corrected for telluric effects with observations of standard stars, this nonetheless introduces a source of systematic error. Similarly, the subtraction of the host continuum also introduces a source of systematic error. However, following \citet{Gezari15}, we minimize this source of uncertainty by performing our host subtraction with a late-time observation of the host with a similar observational setup as was used to observe the TDE flare. Our overall flux normalization is also a source of systematic error, but the typical uncertainty in our photometry is $\lesssim$~0.02 mag. More importantly, variations in the seeing and slit configuration directly affect the spatial regions probed by our observations. These effects are minimized by the fact that our slit alignment is largely consistent between epochs, and are likely to be insignificant relative to the overall evolution of the \halpha\ luminosity, which spans more than two orders of magnitude. Finally, as noted in \citet{Gezari15}, the \ion{He}{ii} line ($n = 6 \rightarrow4$) at 6560\AA\ likely introduces some bias to our measurement of \halpha. This is particularly true $\sim$~130 days after discovery, when the \heii\ line is comparable in strength to \halpha. However, the \ion{He}{ii} line at 6560\AA\ is an order of magnitude weaker than \heii\ in photoionized gas \citep{Osterbrock89}, so even when the relative strength of \ion{He}{ii} is at its highest, there is likely to be only marginal contamination of \halpha.

\subsection{Photometry}
\label{sec:phot}
\begin{figure}
\centering{\includegraphics[scale=1.,width=0.45\textwidth,trim=0.pt 0.pt 0.pt 0.pt,clip]{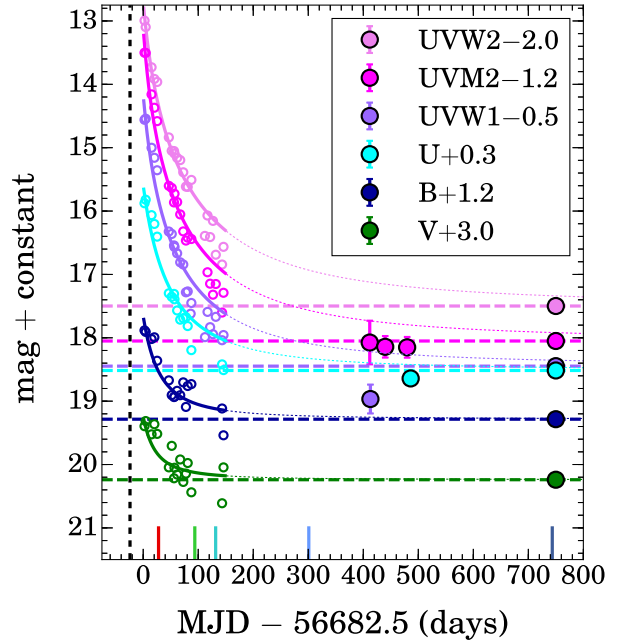}}
\caption{Non-host-subtracted \SWIFT\ UVOT photometry of \asae. The open circles denote data previously published in \citet{Holoien14}; the filled circles show the data presented in this work. The solid curves show power-law fits to the previously published data. The horizontal dashed lines show the host magnitudes measured from the late-time observations, and the vertical dashed line denotes the date of the ASAS-SN pre-discovery non-detection of \asae\ 24.0 days before discovery. The vertical marks along the $x$-axis show the dates of our spectroscopic observations.}
\label{fig:swift}
\end{figure}

In Figure~\ref{fig:swift} we show the UVOT photometric evolution of \asae. The small open circles show data previously presented in \citet{Holoien14}; the filled circles show the data presented in this work. The solid lines show power-law fits to the early data points. We assume a simple model for the host subtracted flux $f_{\lambda} \propto (t-t_0)^{-5/3}$ \citep{Phinney89,Evans89}. We allow $t_0$ to vary within the 24 day window constrained by our pre-discovery ASAS-SN non-detection. The horizontal dashed lines show the host magnitudes measured from the late-time observations, which agree well with the host magnitudes estimated from the SED modeling in \citet{Holoien14}.

The agreement between our late-time photometric points and the host magnitude estimates based on optical and near-IR archival photometry from \citet{Holoien14} provides further evidence that the emission from the TDE has faded completely by $\sim$~750 days. Our observations between 400 and 500 days are consistent with host magnitudes, but these observations are less constraining because they only include only three of the six UVOT filters, and are not as deep as those taken at later times. Given the simplicity of the model, the power-law fits agree reasonably well with the early observations. We note that, as we showed in Figure~\ref{fig:lineEvol}, the power-law models tend to over-predict the observed brightness after $\sim$~100 days. In \citet{Holoien14}, we showed that these observations can be described with a more rapidly declining exponential model. Our non-detection of excess flux at $\sim$~400 days provides additional support for an evolution timescale that is faster than the canonical $t^{-5/3}$ value. We note that, particularly at late-times, the accretion flow is likely to be highly sub-Eddington. In general, the photometric evolution agrees well with our detection of low level \halpha\ emission at 300 days and the lack of optical emission in excess of the host at $\sim$~750 days.

We do not detect X-ray emission from the TDE to a 3-sigma upper limit of $4.1\times10^{-4}$~counts~s$^{-1}$. Assuming a power law spectrum with $\Gamma=2$ and Galactic \ion{H}{1} column density \citep{Kalberla05}, as was done in \citet{Holoien14}, yields an upper limit on the X-ray flux of $f_X~\leq~1.5\times10^{-14}$~ergs~cm$^2$~s$^{-1}$. This corresponds to a limit of $L_X \leq 6.7\times10^{40}$~ergs~s$^{-1}$ ($1.7\times10^7$~L$_{\odot}$) on the average X-ray luminosity at the distance of \asae, which is a slightly tighter constraint than was presented in \citet{Holoien14}.


\section{Conclusions}
\label{sec:conclusions}

\begin{figure}
\centering{\includegraphics[scale=1.,width=0.45\textwidth,trim=0.pt 0.pt 0.pt 0.pt,clip]{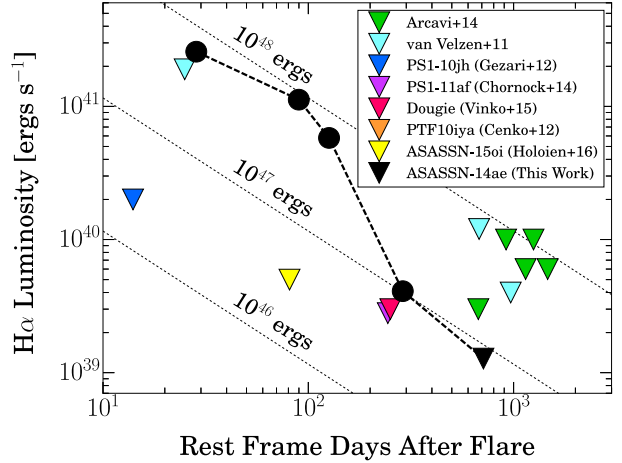}}
\caption{Estimated limits on late-time \halpha\ emission from optical TDE candidates. The late-time spectra of most of these objects are simply assumed to be host dominated and lack upper limit estimates. For these objects, we assume an upper limit of 2\AA\ for the \halpha\ emission, and compute the luminosity based on the approximate continuum and distances to the hosts available for each TDE. Even with the optimistic limits for the higher redshift TDEs, the proximity of \asae\ allows us to place an unprecedented limit on the late-time \halpha\ emission. The dotted lines show lines of constant energy.}
\label{fig:ensemble}
\end{figure}

We have presented late-time optical follow-up spectra of \asae\ with LBT/MODS1 alongside a pre-discovery ASAS-SN non-detection and late-time UVOT and XRT observations from \SWIFT. Our observations span from $\sim$~20 days before to $\sim$~750 days after discovery, and is the first study to follow the evolution of a TDE from a flare dominated state to a host dominated state and place such stringent limits on the late-time emission. If \asae\ is representative of TDEs in general, our findings suggest that optical emission can vanish on a timescale as short as $\sim$~1 yr. This is in agreement with previous studies of some TDE candidates \citep[e.g.][]{Cenko12,Vinko15,Holoien16_15oi}, but is not ubiquitous among TDEs (Brown et al. 2016, in preparation).

It is interesting to consider how our results for \asae\ compare to the larger population of optical TDE candidates. In particular, our new stringent limit on the \halpha\ emission of \asae\ may provide insight into what drives TDE emission at late times. For instance, \citet{Roth15} show that the continuum emission and line strengths are likely to be highly dependent on the physical conditions of the reprocessing envelope. Our observations require that any physical model for \asae\ satisfy our observational constraints on the luminosity and equivalent width evolution of the \halpha\ emission, as well as our non-detection limits for late-time optical emission. 

However, \asae\ is only one in a diverse assortment of TDE candidates \citep{vanVelzen11,Gezari12,Cenko12,Chornock14,Arcavi14,Vinko15,Holoien16_15oi}. In Figure~\ref{fig:ensemble} we show how our limit on \halpha\ emission from \asae\ compares to approximate limits for other optical TDE candidates. Many of the transients in Figure~\ref{fig:ensemble} lack formal upper limit estimates of the \halpha\ emission and simply assume that their late-time spectra are host dominated. For these cases (TDE1; \citealp{vanVelzen11}, the five TDE candidates from \citealp{Arcavi14}, Dougie; \citealp{Vinko15}, and ASASSN-15oi; \citealp{Holoien16_15oi}) we assume an upper limit on the \halpha\ equivalent width of 2\AA\ and compute a limit on the \halpha\ luminosity based on an estimate of the host continuum. The assumption of 2\AA\ is relatively generous, given that, for a line width of $\sim$~3000 km s$^{-1}$, a 2\AA\ equivalent width corresponds to a flux excess of only a few percent relative to the continuum. This measurement can be challenging for higher redshift objects, especially if they are lacking archival spectra. For TDE2 \citep{vanVelzen11}, PS1-11af \citep{Chornock14}, and PTF10iya \citep{Cenko12}, we adopt the authors' measurements of \halpha\ emission as fiducial upper limits on the \halpha\ emission arising from the TDE. We emphasize that these limits are unlikely to be accurate to better than a factor of a few. However, they demonstrate how the proximity of \asae\, coupled with our dedicated follow-up effort, allows us to place an unusually stringent limit on late-time \halpha\ luminosity with a relatively modest allocation of telescope resources. As the number of TDEs with extensive optical follow-up grows, characterizing the distribution of flare lifetimes will be crucial for understanding TDE demographics \citep{Kochanek16_demo}.

The characterization of host galaxies is also critical for understanding the physical conditions that produce TDEs. The host of \asae, like the hosts of several other TDE candidates, shows the strong Balmer absorption features associated with a relatively recent ($\sim$~2 Gyr) burst of star formation. In fact, Balmer-strong galaxies are significantly overrepresented in samples of TDE host galaxies \citep{Arcavi14,French16}. Coupled with the TDE rate estimate in E+A galaxies from \citet{French16} ($\sim 10^{-3}$ yr$^{-1}$ galaxy$^{-1}$), our results suggest that on the order of one out of every 1000 E+A galaxies should show evidence for emission from a recent TDE. A cursory search of previously studied E+A galaxies that show time variable \halpha\ and/or \heii\ emission features could yield several new TDE candidates. Significant changes in the equivalent width or relative strength of these lines would be consistent with accretion events observed in AGN \citep[e.g.][]{Peterson86}, and in the absence of previous nuclear activity, would be strongly indicative of a recent TDE. 

Similarly, E+A galaxies are rare and make up only $\sim$~1\% of SDSS galaxies \citep{Quintero04,French16,Kochanek16_demo}. However, a spectroscopic survey of many thousands of galaxies will likely include some number of E+A galaxies. For instance, MaNGA \citep{Bundy15} will eventually obtain spatially resolved spectroscopy of 10$^4$ galaxies, approximately 100 of which will resemble the Balmer-strong galaxies from \citet{French16}. Most observations of MaNGA galaxies will rarely span more than one night \citep{Bundy15}, but they will all have prior SDSS spectra. Thus E+A galaxies with peculiar nuclear \halpha\ and/or \heii\ emission features could potentially be explained with a recent TDE. While a more complete understanding of the relationship between host galaxy properties and TDE rates is necessary, TDE signatures in excess of 1 out of every 1000 E+A galaxies would provide strong evidence for further peculiarity of TDE hosts, such as complex debris streams, sharply peaked TDE rates, or SMBH binaries \citep[e.g.][]{Wegg11,Li15,Ricarte15}. Spectroscopic monitoring of TDEs during the years following discovery will continue to be crucial for constraining the nuclear properties of TDEs as well as their hosts.

\section*{Acknowledgements}
We thank the referee for a constructive report. 

BJS is supported by NASA through Hubble Fellowship grant HF-51348.001 awarded by the Space Telescope Science Institute, which is operated by the Association of Universities for Research in Astronomy, Inc., for NASA, under contract NAS 5-26555. 

TW-SH is supported by the DOE Computational Science Graduate Fellowship, grant number DE-FG02-97ER25308.

CSK is supported by NSF grants AST-1515876 and AST-1515927.

This paper used data obtained with the MODS spectrographs built with funding from NSF grant AST-9987045 and the NSF Telescope System Instrumentation Program (TSIP), with additional funds from the Ohio Board of Regents and the Ohio State University Office of Research.

Based on data acquired using the Large Binocular Telescope (LBT). The LBT is an international collaboration among institutions in the United States, Italy, and Germany. LBT Corporation partners are: The University of Arizona on behalf of the Arizona university system; Istituto Nazionale di Astrofisica, Italy; LBT Beteiligungsgesellschaft, Germany, representing the Max-Planck Society, the Astrophysical Institute Potsdam, and Heidelberg University; The Ohio State University, and The Research Corporation, on behalf of The University of Notre Dame, University of Minnesota and University of Virginia

\bibliography{late14ae}
\bsp	

\label{lastpage}

\end{document}

%% file: tab1.tex
\begin{table*}
\caption{Observations. \label{tab:tab1}}

\begin{minipage}{\textwidth}
\begin{tabular}{@{}lrrrcrcrrrrrrrrrrrrr}
\hline
\hline

{} & {} & \multicolumn{1}{c}{HJD} & \multicolumn{1}{c}{P.A.$^{b}$} & \multicolumn{1}{c}{Par. P.A.$^{c}$} & {} & \multicolumn{1}{c}{Flux$^{d}$} & \multicolumn{1}{c}{Seeing} & \multicolumn{1}{c}{Exposure} & \multicolumn{1}{c}{$r'^{e}$}\\
\multicolumn{1}{c}{UT Date} & \multicolumn{1}{c}{Day$^{a}$} & \multicolumn{1}{c}{$-2,400,000$} & \multicolumn{1}{c}{[deg]} & \multicolumn{1}{c}{[deg]} & \multicolumn{1}{c}{Airmass$^{c}$} & \multicolumn{1}{c}{Standard} & \multicolumn{1}{c}{[arcsec]} & \multicolumn{1}{c}{[s]$\times$n} & \multicolumn{1}{c}{[mag]}\\

\hline

2014 Feb 24.20 & 29.69  & 56712.20  &  $90.0$ & $-76$ to $-77$ & 1.33 -- 1.39 & Feige 34  & 1.1 &  300.0$\times$3 & $16.78 \pm 0.03$\\

2014 Apr 29.29 & 93.78 & 56776.29 & $90.0$ & $80$ to $77$ & 1.20 -- 1.33 & HZ 44 & 1.1 & 1050.0$\times$3 & $16.84 \pm 0.01$\\

2014 Jun 06.18 & 131.67 & 56814.18  & $90.0$ & $82$ to $78$ & 1.18 -- 1.30 & GD 153 & 0.75 & 1050.0$\times$3 & $16.90 \pm 0.02$\\

2014 Nov 22.49 & 300.98 & 56983.49 & $105.0$ & $-73$ to $-83$ & 1.15 -- 1.60 & Feige 34 & 0.9 & 1200.0$\times$6 & $16.94 \pm 0.02$\\

2016 Feb 08.33 & 743.48 & 57426.33 & $95.0$ & $\hphantom{.}-90$ to $-114$ & 1.04 -- 1.00 & G191-B2B & 1.1 & 600.0$\times$3 & $16.94 \pm 0.02$\\

\hline
\end{tabular}

\medskip

$^{a}$ Days since discovery ($t_{disc} = 56682.51$; \citealp{Prieto14}).\\
$^{b}$ Position angle of the spectrograph slit.\\
$^{c}$ Par. P.A. and airmass give the range of parallactic angles and airmasses for each observation, respectively.\\
$^{d}$ Standard stars were observed on the same night as science observations.\\
$^{e}$ $r^{\prime}$ magnitude is derived from the acquisition images.\\

\end{minipage}
\end{table*}

%% file: tab2.tex
\begin{table*} 
\caption{Measurements of \halpha\ Profile Properties \label{tab:tab2}} 

\begin{minipage}{\textwidth} 
\begin{tabular}{@{}lrrrrrr} 
\hline 
\hline 
{} & {} & \multicolumn{1}{c}{$\Delta v^{a}$} & \multicolumn{1}{c}{FWHM$^{b}$} & \multicolumn{1}{c}{\halpha\ Flux} & \multicolumn{1}{c}{\halpha\ Equivalent Width} & \multicolumn{1}{c}{\halpha\ Luminosity} \\ 
\multicolumn{1}{c}{UT Date} & \multicolumn{1}{c}{Day} & \multicolumn{1}{c}{$10^3$ km s$^{-1}$} & \multicolumn{1}{c}{$10^3$ km s$^{-1}$} & \multicolumn{1}{c}{[$10^{-14}$ ergs s$^{-1}$ cm$^{-2}$]} & \multicolumn{1}{c}{[\AA]} & {[$10^{40}$ ergs s$^{-1}$]} \\ 
\hline 

{}2014 Feb 24.20 & 29.69  & $  1.81\hphantom{oooo} $ & $ 13.34\hphantom{ooo} $ & $  5.78 \pm  0.07\hphantom{ooooo} $ & $ 662.5 \pm   5.2\hphantom{ooooo} $ & $25.76 \pm   0.31\hphantom{o} $ \\ 
{}2014 Apr 29.29 & 93.78 & $  1.15\hphantom{oooo} $ & $  7.79\hphantom{ooo} $ & $  2.52 \pm  0.02\hphantom{ooooo} $ & $ 596.3 \pm   3.0\hphantom{ooooo} $ & $11.23 \pm   0.17\hphantom{o} $ \\ 
{}2014 Jun 6.18 & 131.67 & $  0.37\hphantom{oooo} $ & $  4.86\hphantom{ooo} $ & $  1.30 \pm  0.03\hphantom{ooooo} $ & $ 707.4 \pm  13.0\hphantom{ooooo} $ & $ 5.80 \pm   0.12\hphantom{o} $ \\ 
{}2014 Nov 22.49 & 300.98 & $ -0.20\hphantom{oooo} $ & $  2.46\hphantom{ooo} $ & $  0.09 \pm  0.01\hphantom{ooooo} $ & $ 725.7 \pm  16.1\hphantom{ooooo} $ & $ 0.41 \pm   0.03\hphantom{o} $ \\ 
{}2016 Feb 08.33 & 743.48 & \ldots\hphantom{oooo} & \ldots\hphantom{ooo} & $< 0.03\hphantom{ooooo} $ & \ldots\hphantom{oooooooo}  & $ <  0.13\hphantom{o} $ \\ 
\hline 
\end{tabular} 

\medskip 

$^{a}$ Relative to systemic velocity of the host at $z=0.0436$.\\  
$^{b}$ Estimated from model fit to \halpha\ profile.\\  

\end{minipage} 
\end{table*}

%% file: late14ae.bbl
\begin{thebibliography}{55}
\expandafter\ifx\csname natexlab\endcsname\relax\def\natexlab#1{#1}\fi

\bibitem[{{Abazajian} {et~al}\mbox{.}(2009){Abazajian}, {Adelman-McCarthy},
  {Ag{\"u}eros}, {Allam}, {Allende Prieto}, {An}, {Anderson}, {Anderson},
  {Annis}, {Bahcall}, \& et~al.}]{Abazajian09}
{Abazajian} K.~N. {et~al.}, 2009, \apjs, 182, 543

\bibitem[{{Alard}(2000)}]{Alard00}
{Alard} C., 2000, \aaps, 144, 363

\bibitem[{{Alard} \& {Lupton}(1998)}]{Alard98}
{Alard} C., {Lupton} R.~H., 1998, \apj, 503, 325

\bibitem[{{Arcavi} {et~al}\mbox{.}(2014){Arcavi}, {Gal-Yam}, {Sullivan}, {Pan},
  {Cenko}, {Horesh}, {Ofek}, {De Cia}, {Yan}, {Yang}, {Howell}, {Tal},
  {Kulkarni}, {Tendulkar}, {Tang}, {Xu}, {Sternberg}, {Cohen}, {Bloom},
  {Nugent}, {Kasliwal}, {Perley}, {Quimby}, {Miller}, {Theissen}, \&
  {Laher}}]{Arcavi14}
{Arcavi} I. {et~al.}, 2014, \apj, 793, 38

\bibitem[{{Bohlin}, {Colina} \& {Finley}(1995){Bohlin}, {Colina}, \&
  {Finley}}]{Bohlin95}
{Bohlin} R.~C., {Colina} L., {Finley} D.~S., 1995, \aj, 110, 1316

\bibitem[{{Breeveld} {et~al}\mbox{.}(2010){Breeveld}, {Curran}, {Hoversten},
  {Koch}, {Landsman}, {Marshall}, {Page}, {Poole}, {Roming}, {Smith}, {Still},
  {Yershov}, {Blustin}, {Brown}, {Gronwall}, {Holland}, {Kuin}, {McGowan},
  {Rosen}, {Boyd}, {Broos}, {Carter}, {Chester}, {Hancock}, {Huckle}, {Immler},
  {Ivanushkina}, {Kennedy}, {Mason}, {Morgan}, {Oates}, {de Pasquale},
  {Schady}, {Siegel}, \& {vanden Berk}}]{Breeveld10}
{Breeveld} A.~A. {et~al.}, 2010, \mnras, 406, 1687

\bibitem[{{Bundy} {et~al}\mbox{.}(2015){Bundy}, {Bershady}, {Law}, {Yan},
  {Drory}, {MacDonald}, {Wake}, {Cherinka}, {S{\'a}nchez-Gallego}, {Weijmans},
  {Thomas}, {Tremonti}, {Masters}, {Coccato}, {Diamond-Stanic},
  {Arag{\'o}n-Salamanca}, {Avila-Reese}, {Badenes}, {Falc{\'o}n-Barroso},
  {Belfiore}, {Bizyaev}, {Blanc}, {Bland-Hawthorn}, {Blanton}, {Brownstein},
  {Byler}, {Cappellari}, {Conroy}, {Dutton}, {Emsellem}, {Etherington},
  {Frinchaboy}, {Fu}, {Gunn}, {Harding}, {Johnston}, {Kauffmann}, {Kinemuchi},
  {Klaene}, {Knapen}, {Leauthaud}, {Li}, {Lin}, {Maiolino}, {Malanushenko},
  {Malanushenko}, {Mao}, {Maraston}, {McDermid}, {Merrifield}, {Nichol},
  {Oravetz}, {Pan}, {Parejko}, {Sanchez}, {Schlegel}, {Simmons}, {Steele},
  {Steinmetz}, {Thanjavur}, {Thompson}, {Tinker}, {van den Bosch}, {Westfall},
  {Wilkinson}, {Wright}, {Xiao}, \& {Zhang}}]{Bundy15}
{Bundy} K. {et~al.}, 2015, \apj, 798, 7

\bibitem[{{Burrows} {et~al}\mbox{.}(2005){Burrows}, {Hill}, {Nousek}, {Kennea},
  {Wells}, {Osborne}, {Abbey}, {Beardmore}, {Mukerjee}, {Short}, {Chincarini},
  {Campana}, {Citterio}, {Moretti}, {Pagani}, {Tagliaferri}, {Giommi},
  {Capalbi}, {Tamburelli}, {Angelini}, {Cusumano}, {Br{\"a}uninger}, {Burkert},
  \& {Hartner}}]{Burrows05}
{Burrows} D.~N. {et~al.}, 2005, \ssr, 120, 165

\bibitem[{{Cenko} {et~al}\mbox{.}(2012){Cenko}, {Bloom}, {Kulkarni}, {Strubbe},
  {Miller}, {Butler}, {Quimby}, {Gal-Yam}, {Ofek}, {Quataert}, {Bildsten},
  {Poznanski}, {Perley}, {Morgan}, {Filippenko}, {Frail}, {Arcavi}, {Ben-Ami},
  {Cucchiara}, {Fassnacht}, {Green}, {Hook}, {Howell}, {Lagattuta}, {Law},
  {Kasliwal}, {Nugent}, {Silverman}, {Sullivan}, {Tendulkar}, \&
  {Yaron}}]{Cenko12}
{Cenko} S.~B. {et~al.}, 2012, \mnras, 420, 2684

\bibitem[{{Chornock} {et~al}\mbox{.}(2014){Chornock}, {Berger}, {Gezari},
  {Zauderer}, {Rest}, {Chomiuk}, {Kamble}, {Soderberg}, {Czekala}, {Dittmann},
  {Drout}, {Foley}, {Fong}, {Huber}, {Kirshner}, {Lawrence}, {Lunnan},
  {Marion}, {Narayan}, {Riess}, {Roth}, {Sanders}, {Scolnic}, {Smartt},
  {Smith}, {Stubbs}, {Tonry}, {Burgett}, {Chambers}, {Flewelling}, {Hodapp},
  {Kaiser}, {Magnier}, {Martin}, {Neill}, {Price}, \& {Wainscoat}}]{Chornock14}
{Chornock} R. {et~al.}, 2014, \apj, 780, 44

\bibitem[{{Denney} {et~al}\mbox{.}(2009){Denney}, {Peterson}, {Dietrich},
  {Vestergaard}, \& {Bentz}}]{Denney09}
{Denney} K.~D., {Peterson} B.~M., {Dietrich} M., {Vestergaard} M., {Bentz}
  M.~C., 2009, \apj, 692, 246

\bibitem[{{Evans} \& {Kochanek}(1989)}]{Evans89}
{Evans} C.~R., {Kochanek} C.~S., 1989, \apjl, 346, L13

\bibitem[{{French}, {Arcavi} \& {Zabludoff}(2016){French}, {Arcavi}, \&
  {Zabludoff}}]{French16}
{French} K.~D., {Arcavi} I., {Zabludoff} A., 2016, \apjl, 818, L21

\bibitem[{{Gaskell} \& {Rojas Lobos}(2014)}]{Gaskell14}
{Gaskell} C.~M., {Rojas Lobos} P.~A., 2014, \mnras, 438, L36

\bibitem[{{Gezari} {et~al}\mbox{.}(2015){Gezari}, {Chornock}, {Lawrence},
  {Rest}, {Jones}, {Berger}, {Challis}, \& {Narayan}}]{Gezari15}
{Gezari} S., {Chornock} R., {Lawrence} A., {Rest} A., {Jones} D.~O., {Berger}
  E., {Challis} P.~M., {Narayan} G., 2015, \apjl, 815, L5

\bibitem[{{Gezari} {et~al}\mbox{.}(2012){Gezari}, {Chornock}, {Rest}, {Huber},
  {Forster}, {Berger}, {Challis}, {Neill}, {Martin}, {Heckman}, {Lawrence},
  {Norman}, {Narayan}, {Foley}, {Marion}, {Scolnic}, {Chomiuk}, {Soderberg},
  {Smith}, {Kirshner}, {Riess}, {Smartt}, {Stubbs}, {Tonry}, {Wood-Vasey},
  {Burgett}, {Chambers}, {Grav}, {Heasley}, {Kaiser}, {Kudritzki}, {Magnier},
  {Morgan}, \& {Price}}]{Gezari12}
{Gezari} S. {et~al.}, 2012, \nat, 485, 217

\bibitem[{{Guillochon}, {Manukian} \& {Ramirez-Ruiz}(2014){Guillochon},
  {Manukian}, \& {Ramirez-Ruiz}}]{Guillochon14}
{Guillochon} J., {Manukian} H., {Ramirez-Ruiz} E., 2014, \apj, 783, 23

\bibitem[{{Guillochon} \& {Ramirez-Ruiz}(2013)}]{Guillochon13}
{Guillochon} J., {Ramirez-Ruiz} E., 2013, \apj, 767, 25

\bibitem[{{Henden} {et~al}\mbox{.}(2015){Henden}, {Levine}, {Terrell}, \&
  {Welch}}]{Henden15}
{Henden} A.~A., {Levine} S., {Terrell} D., {Welch} D.~L., 2015, in American
  Astronomical Society Meeting Abstracts, Vol. 225, American Astronomical
  Society Meeting Abstracts, p. 336.16

\bibitem[{{Hill} {et~al}\mbox{.}(2004){Hill}, {Burrows}, {Nousek}, {Abbey},
  {Ambrosi}, {Br{\"a}uninger}, {Burkert}, {Campana}, {Cheruvu}, {Cusumano},
  {Freyberg}, {Hartner}, {Klar}, {Mangels}, {Moretti}, {Mori}, {Morris},
  {Short}, {Tagliaferri}, {Watson}, {Wood}, \& {Wells}}]{Hill04}
{Hill} J.~E. {et~al.}, 2004, in \procspie, Vol. 5165, X-Ray and Gamma-Ray
  Instrumentation for Astronomy XIII, {Flanagan} K.~A., {Siegmund} O.~H.~W.,
  eds., pp. 217--231

\bibitem[{{Holoien} {et~al}\mbox{.}(2016{\natexlab{a}}){Holoien}, {Kochanek},
  {Prieto}, {Grupe}, {Chen}, {Godoy-Rivera}, {Stanek}, {Shappee}, {Dong},
  {Brown}, {Basu}, {Beacom}, {Bersier}, {Brimacombe}, {Carlson}, {Falco},
  {Johnston}, {Madore}, {Pojmanski}, \& {Seibert}}]{Holoien16_15oi}
{Holoien} T.~W.-S. {et~al.}, 2016{\natexlab{a}}, ArXiv e-prints

\bibitem[{{Holoien} {et~al}\mbox{.}(2016{\natexlab{b}}){Holoien}, {Kochanek},
  {Prieto}, {Stanek}, {Dong}, {Shappee}, {Grupe}, {Brown}, {Basu}, {Beacom},
  {Bersier}, {Brimacombe}, {Danilet}, {Falco}, {Guo}, {Jose}, {Herczeg},
  {Long}, {Pojmanski}, {Simonian}, {Szczygie{\l}}, {Thompson}, {Thorstensen},
  {Wagner}, \& {Wo{\'z}niak}}]{Holoien16_14li}
{Holoien} T.~W.-S. {et~al.}, 2016{\natexlab{b}}, \mnras, 455, 2918

\bibitem[{{Holoien} {et~al}\mbox{.}(2014){Holoien}, {Prieto}, {Bersier},
  {Kochanek}, {Stanek}, {Shappee}, {Grupe}, {Basu}, {Beacom}, {Brimacombe},
  {Brown}, {Davis}, {Jencson}, {Pojmanski}, \& {Szczygie{\l}}}]{Holoien14}
{Holoien} T.~W.-S. {et~al.}, 2014, \mnras, 445, 3263

\bibitem[{{Kalberla} {et~al}\mbox{.}(2005){Kalberla}, {Burton}, {Hartmann},
  {Arnal}, {Bajaja}, {Morras}, \& {P{\"o}ppel}}]{Kalberla05}
{Kalberla} P.~M.~W., {Burton} W.~B., {Hartmann} D., {Arnal} E.~M., {Bajaja} E.,
  {Morras} R., {P{\"o}ppel} W.~G.~L., 2005, \aap, 440, 775

\bibitem[{{Kochanek}(1994)}]{Kochanek94}
{Kochanek} C.~S., 1994, \apj, 422, 508

\bibitem[{{Kochanek}(2016{\natexlab{a}})}]{Kochanek16_stellar}
{Kochanek} C.~S., 2016{\natexlab{a}}, \mnras

\bibitem[{{Kochanek}(2016{\natexlab{b}})}]{Kochanek16_demo}
{Kochanek} C.~S., 2016{\natexlab{b}}, ArXiv e-prints

\bibitem[{{Li} {et~al}\mbox{.}(2015){Li}, {Liu}, {Berczik}, \&
  {Spurzem}}]{Li15}
{Li} S., {Liu} F.~K., {Berczik} P., {Spurzem} R., 2015, ArXiv e-prints

\bibitem[{{Lodato} \& {Rossi}(2011)}]{Lodato11}
{Lodato} G., {Rossi} E.~M., 2011, \mnras, 410, 359

\bibitem[{{MacLeod}, {Guillochon} \& {Ramirez-Ruiz}(2012){MacLeod},
  {Guillochon}, \& {Ramirez-Ruiz}}]{MacLeod12}
{MacLeod} M., {Guillochon} J., {Ramirez-Ruiz} E., 2012, \apj, 757, 134

\bibitem[{{Magorrian} \& {Tremaine}(1999)}]{Magorrian99}
{Magorrian} J., {Tremaine} S., 1999, \mnras, 309, 447

\bibitem[{{McGill} {et~al}\mbox{.}(2008){McGill}, {Woo}, {Treu}, \&
  {Malkan}}]{McGill08}
{McGill} K.~L., {Woo} J.-H., {Treu} T., {Malkan} M.~A., 2008, \apj, 673, 703

\bibitem[{{Metzger} \& {Stone}(2015)}]{Metzger15}
{Metzger} B.~D., {Stone} N.~C., 2015, ArXiv e-prints

\bibitem[{{Oke}(1990)}]{Oke90}
{Oke} J.~B., 1990, \aj, 99, 1621

\bibitem[{{Osterbrock}(1989)}]{Osterbrock89}
{Osterbrock} D.~E., 1989, {Astrophysics of gaseous nebulae and active galactic
  nuclei}

\bibitem[{{Peterson} \& {Ferland}(1986)}]{Peterson86}
{Peterson} B.~M., {Ferland} G.~J., 1986, \nat, 324, 345

\bibitem[{{Phinney}(1989)}]{Phinney89}
{Phinney} E.~S., 1989, in IAU Symposium, Vol. 136, The Center of the Galaxy,
  {Morris} M., ed., p. 543

\bibitem[{{Piran} {et~al}\mbox{.}(2015){Piran}, {Svirski}, {Krolik}, {Cheng},
  \& {Shiokawa}}]{Piran15}
{Piran} T., {Svirski} G., {Krolik} J., {Cheng} R.~M., {Shiokawa} H., 2015,
  \apj, 806, 164

\bibitem[{{Pogge} {et~al}\mbox{.}(2010){Pogge}, {Atwood}, {Brewer}, {Byard},
  {Derwent}, {Gonzalez}, {Martini}, {Mason}, {O'Brien}, {Osmer}, {Pappalardo},
  {Steinbrecher}, {Teiga}, \& {Zhelem}}]{Pogge10}
{Pogge} R.~W. {et~al.}, 2010, in Society of Photo-Optical Instrumentation
  Engineers (SPIE) Conference Series, Vol. 7735, Society of Photo-Optical
  Instrumentation Engineers (SPIE) Conference Series

\bibitem[{{Poole} {et~al}\mbox{.}(2008){Poole}, {Breeveld}, {Page}, {Landsman},
  {Holland}, {Roming}, {Kuin}, {Brown}, {Gronwall}, {Hunsberger}, {Koch},
  {Mason}, {Schady}, {vanden Berk}, {Blustin}, {Boyd}, {Broos}, {Carter},
  {Chester}, {Cucchiara}, {Hancock}, {Huckle}, {Immler}, {Ivanushkina},
  {Kennedy}, {Marshall}, {Morgan}, {Pandey}, {de Pasquale}, {Smith}, \&
  {Still}}]{Poole08}
{Poole} T.~S. {et~al.}, 2008, \mnras, 383, 627

\bibitem[{{Prieto} {et~al}\mbox{.}(2014){Prieto}, {Bersier}, {Holoien},
  {Shappee}, {Stanek}, {Kochanek}, {Jencson}, {Beacom}, {Szczygiel},
  {Pojmanski}, \& {Brimacombe}}]{Prieto14}
{Prieto} J.~L. {et~al.}, 2014, The Astronomer's Telegram, 5831, 1

\bibitem[{{Quintero} {et~al}\mbox{.}(2004){Quintero}, {Hogg}, {Blanton},
  {Schlegel}, {Eisenstein}, {Gunn}, {Brinkmann}, {Fukugita}, {Glazebrook}, \&
  {Goto}}]{Quintero04}
{Quintero} A.~D. {et~al.}, 2004, \apj, 602, 190

\bibitem[{{Rees}(1988)}]{Rees88}
{Rees} M.~J., 1988, \nat, 333, 523

\bibitem[{{Ricarte} {et~al}\mbox{.}(2015){Ricarte}, {Natarajan}, {Dai}, \&
  {Coppi}}]{Ricarte15}
{Ricarte} A., {Natarajan} P., {Dai} L., {Coppi} P., 2015, ArXiv e-prints

\bibitem[{{Roth} {et~al}\mbox{.}(2015){Roth}, {Kasen}, {Guillochon}, \&
  {Ramirez-Ruiz}}]{Roth15}
{Roth} N., {Kasen} D., {Guillochon} J., {Ramirez-Ruiz} E., 2015, ArXiv e-prints

\bibitem[{{Shappee} {et~al}\mbox{.}(2014){Shappee}, {Prieto}, {Grupe},
  {Kochanek}, {Stanek}, {De Rosa}, {Mathur}, {Zu}, {Peterson}, {Pogge},
  {Komossa}, {Im}, {Jencson}, {Holoien}, {Basu}, {Beacom}, {Szczygie{\l}},
  {Brimacombe}, {Adams}, {Campillay}, {Choi}, {Contreras}, {Dietrich},
  {Dubberley}, {Elphick}, {Foale}, {Giustini}, {Gonzalez}, {Hawkins}, {Howell},
  {Hsiao}, {Koss}, {Leighly}, {Morrell}, {Mudd}, {Mullins}, {Nugent},
  {Parrent}, {Phillips}, {Pojmanski}, {Rosing}, {Ross}, {Sand}, {Terndrup},
  {Valenti}, {Walker}, \& {Yoon}}]{Shappee14}
{Shappee} B.~J. {et~al.}, 2014, \apj, 788, 48

\bibitem[{{Shappee} {et~al}\mbox{.}(2013){Shappee}, {Stanek}, {Pogge}, \&
  {Garnavich}}]{Shappee13}
{Shappee} B.~J., {Stanek} K.~Z., {Pogge} R.~W., {Garnavich} P.~M., 2013, \apjl,
  762, L5

\bibitem[{{Shiokawa} {et~al}\mbox{.}(2015){Shiokawa}, {Krolik}, {Cheng},
  {Piran}, \& {Noble}}]{Shiokawa15}
{Shiokawa} H., {Krolik} J.~H., {Cheng} R.~M., {Piran} T., {Noble} S.~C., 2015,
  \apj, 804, 85

\bibitem[{{Strubbe} \& {Murray}(2015)}]{Strubbe15}
{Strubbe} L.~E., {Murray} N., 2015, \mnras, 454, 2321

\bibitem[{{Strubbe} \& {Quataert}(2009)}]{Strubbe09}
{Strubbe} L.~E., {Quataert} E., 2009, \mnras, 400, 2070

\bibitem[{{Ulmer}(1999)}]{Ulmer99}
{Ulmer} A., 1999, \apj, 514, 180

\bibitem[{{van Dokkum}(2001)}]{vanDokkum01}
{van Dokkum} P.~G., 2001, \pasp, 113, 1420

\bibitem[{{van Velzen} {et~al}\mbox{.}(2011){van Velzen}, {Farrar}, {Gezari},
  {Morrell}, {Zaritsky}, {{\"O}stman}, {Smith}, {Gelfand}, \&
  {Drake}}]{vanVelzen11}
{van Velzen} S. {et~al.}, 2011, \apj, 741, 73

\bibitem[{{Vink{\'o}} {et~al}\mbox{.}(2015){Vink{\'o}}, {Yuan}, {Quimby},
  {Wheeler}, {Ramirez-Ruiz}, {Guillochon}, {Chatzopoulos}, {Marion}, \&
  {Akerlof}}]{Vinko15}
{Vink{\'o}} J. {et~al.}, 2015, \apj, 798, 12

\bibitem[{{Wegg} \& {Bode}(2011)}]{Wegg11}
{Wegg} C., {Bode} N., 2011, \apjl, 738, L8

\end{thebibliography}
